\documentclass{article}

\usepackage[preprint]{neurips_2026}

\usepackage[utf8]{inputenc} 
\usepackage[T1]{fontenc}    
\usepackage{hyperref}       
\usepackage{url}            
\usepackage{booktabs}       
\usepackage{amsfonts}       
\usepackage{nicefrac}       
\usepackage{microtype}      
\usepackage{xcolor}         
\usepackage{graphicx}

\usepackage{sidecap}
\usepackage{amssymb}
\usepackage{amsmath}

\workshoptitle{Machine Learning and the Physical Sciences Workshop}
\title{Learning to find cosmic filaments with graph neural networks}

\author{%
  Alexis Tastet \\
  Ecole Polytechnique \\
  Rte de Saclay, 91120 Palaiseau, France \\
  \texttt{alexis.tastet@polytechnique.edu} \\
  \And
  Daniela Gal\'arraga-Espinosa \\ 
  Center for Data-Driven Discovery \\
  Kavli IPMU, UTIAS, The University of Tokyo \\
  Kashiwa, Chiba 277-8583, Japan \\
  \texttt{daniela.galarraga@ipmu.jp} \\
  \And
  Leander Thiele \\ 
  Center for Data-Driven Discovery \\
  Kavli IPMU, UTIAS, The University of Tokyo \\
  Kashiwa, Chiba 277-8583, Japan \\
  \texttt{leander.thiele@ipmu.jp} \\
}

\begin{document}

\maketitle

\begin{abstract}
  Inferring the large-scale structure of the Universe from galaxy surveys is a fundamental challenge in cosmology. In particular, cosmic filaments form an extended and geometrically complex network whose underlying structure cannot be reconstructed from discrete galaxy catalogs without uncertainty. We investigate a generative approach to filament finding based on flow matching and graph neural networks. The learned model reproduces key statistical properties of the target filament distribution while generating a distribution over possible filament configurations, demonstrating its ability to model uncertainty. Our work provides a first step towards probabilistic inference of the cosmic web directly from galaxy surveys. More broadly, these results demonstrate the potential of conditional generative models for reconstructing filament-like structures from sparse observational tracers.
\end{abstract}

\section{Introduction}

On megaparsec scales, matter in the Universe is organised into the cosmic web: an interconnected network of clusters, filaments, walls, and voids produced by the gravitational collapse of the primordial density field~\citep{Zeldovich1970, Bond1996}. Cosmic filaments are the elongated structures forming the backbone of this network, connecting high-density nodes and acting as preferential channels along which dark matter, gas, and galaxies are transported. They span a broad range of lengths, densities, and environments, from prominent bridges between galaxy clusters to tenuous structures extending into underdense regions~\citep[e.g.][]{Cautun2014, GalarragaEspinosa2020, GalarragaEspinosa2024_MTNG, Feldbrugge2025}. Despite their visual prominence, there is no unique definition of a cosmic filament. Existing approaches identify them through different representations of the matter distribution: local eigenvalues of the density, tidal, or velocity fields; multiscale Hessian-based morphology; topological skeletons and persistent homology; stochastic geometric models; or graph-based constructions~\citep[e.g.][]{AragonCalvo2007, Sousbie2011, Cautun2013, Libeskind2018, Bonnaire2020Trex, Awad2023}. These methods broadly recover the most prominent structures but can differ substantially for weaker filaments and in their inferred extent, connectivity, and topology~\citep[][]{Libeskind2018}.

Obtaining robust filament catalogues is becoming increasingly important because the cosmic web is moving from a qualitative description of large-scale structure to a quantitative astrophysical and cosmological observable. Filaments regulate the environments through which galaxies acquire matter and can influence gas accretion, star formation, halo and galaxy spins, and the connectivity of galaxies to their surrounding matter distribution~\citep[e.g.][]{Codis2018, Kraljic2018, GaneshaiahVeena2018_CosmicBalletI, GalarragaEspinosa2023}. At the same time, their geometry, abundance, matter content, and associated galaxy statistics contain information complementary to conventional two-point clustering and are beginning to be explored as probes of the growth of structure, dark matter, and gravity~\citep{Ho2018, Bonnaire2022, Feldbrugge2025, Sunseri2025, Armijo2026}. With surveys such as DESI, Euclid, and Rubin-LSST delivering unprecedented three-dimensional maps of the galaxy distribution, the limiting factor is increasingly not the availability of data but our ability to extract filamentary structure consistently, reproducibly, and with well-characterized uncertainty. This makes filament identification a natural setting for modern machine-learning methods capable of learning complex geometric and topological structure directly from cosmological data.

The multi-scale nature of the cosmic web implies that there is no unique universal definition of filaments.
Once we pick a scale, usually determined by the number density of galaxies in a given catalog, it is clear that a good filament definition should involve the total matter density and velocity fields.
As these fields are not directly observable, the mapping from observed galaxies to underlying filaments is probabilistic and should be modeled as such.
Generative neural networks are thus the natural approach to constructing the desired mapping.

We use conditional generative modelling to solve this problem. Specifically, we use stochastic interpolants flow matching~\citep{Albergo2023,Lipman2023} to learn a continuous transport from a simple particle distribution to a distribution concentrated around cosmic filaments. Starting from a set of particles sampled from a simple, galaxy-conditioned base distribution, our model learns a conditional flow that transports these particles towards a filamentary distribution.
A representation of the filaments by discrete points is not ideal but it is a technical choice motivated by the results of a similar method~\cite{Thiele2026}. The resulting model combines the flexibility of generative modeling with geometric and physical inductive biases appropriate for the cosmological setting.

In this initial study, we focus on a controlled problem in which the training targets are filaments identified by the deterministic DisPerSE algorithm~\cite{Sousbie2011} from the galaxy catalogs and not the dark matter field. This method is currently widely used in filament studies. This allows us to develop and evaluate the generative reconstruction method.
Planned improvements are outlined in the conclusions.
Our method avoids the detour of full field reconstruction~\citep{2023arXiv231108558P,2024ApJ...970..174O,2024arXiv241102496K,2024A&A...691A.348W,2025MNRAS.542.1403D}.

\section{Method}

In this initial work on filament finding, we aim to reproduce the behavior of an existing deterministic filament finder.
This choice is motivated by the more straightforward evaluation this training objective affords and thus is ideal for method development which we consider orthogonal to the precise learning problem.


We train on the Quijote Sobol Sequence (BSQ)~\cite{Villaescusa-Navarro2020} which consists of a collection of N-body simulations involving $512^3$ dark matter particles in a $(1 \, h^{-1}\text{Gpc})^3$ box.
Each box has a different random value of the cosmological parameters drawn from prior bounds. This large setup of cosmological parameters allows us to analyze the dependence of our method on a varying cosmology. The redshift is set to $z=0$. 

As the simulations only model the dark matter, we insert galaxies following the Halo Occupation Distribution (HOD)~\cite{Berlind2002,Cooray2002,Wechsler2018}.
For each box, the galaxy insertion is realized several times with different HOD parameters~\cite{Zheng2005}.
The minimal halo mass $M_{min}$ is adjusted so that the galaxy number per box is stabilized around $N_g = 10^5$.
This results in a collection of 120,000 training samples, which is randomly split into training, validation, and test sets in an 80/10/10 proportion. The method is thus protected against any data leakage. The training and validation sets are used to adjust the hyperparameters by minimizing the validation loss on small batches of training. 

The target filaments are constructed from the galaxy catalogs with DisPerSE~\cite{Sousbie2011,Sousbie2013}.
As we are computationally limited, we discard filaments shorter than $30 \, h^{-1}\text{Mpc}$.


We emphasize that the targets are computed from the galaxies. The final task would be to learn the filaments constructed from the CDM field. Nevertheless, as we do not yet know how to optimally identify filaments, we restrict ourselves to this proxy task with deterministic structure meant to develop the machine learning method.
  

In order to generate filaments from a given galaxy catalog, we use flow matching~\cite{Lipman2023,Liu2022} which is a limit of stochastic interpolants~\cite{Albergo2023,Chen2024}.
The idea is to match the target density $\nu_1$ with an intermediate density $\nu_t$ with $t \in [0,1]$, such that $\nu_0$ is simple.
After sampling $N_p$ test particles from $\nu_0$, the network makes them evolve with a flow $v_t$.
This technique has already been used for void finding~\cite{Thiele2026}.
Related applications can be found in, e.g., Refs.~\citep{2024PhRvD.109l3531C,2023arXiv230612327W}.

\begin{figure}[h!]
    \centering
    \begin{minipage}{.6\textwidth}
        \includegraphics[width=1.\linewidth]{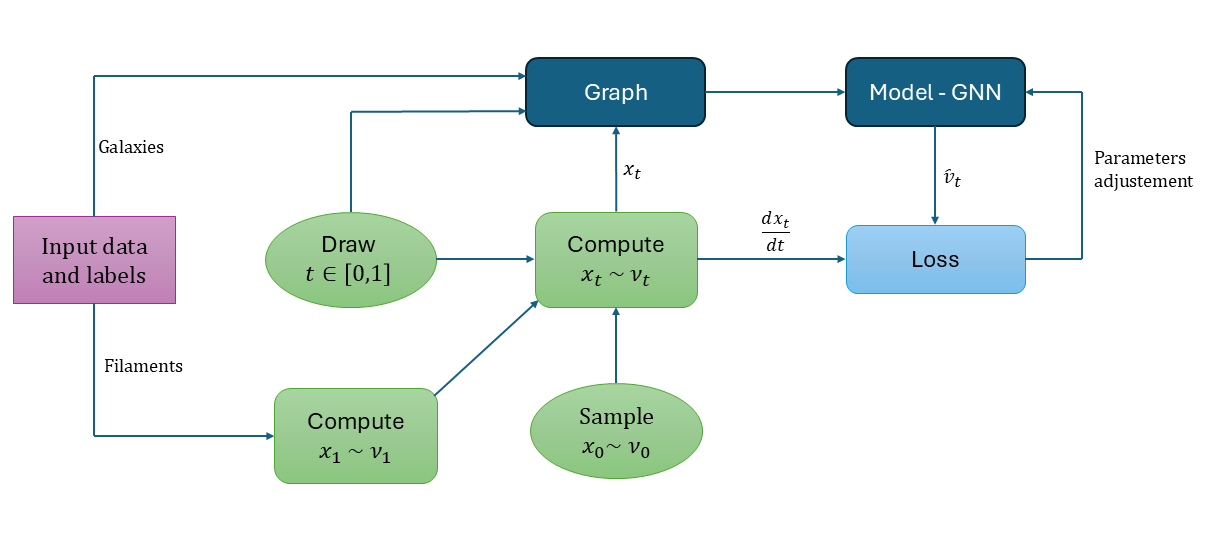}
    \end{minipage}
    \caption{Logigram of the training method. As an input, the galaxies are associated with a label which corresponds to filaments computed by DisPerSE.}\label{fig:logigram}
\end{figure}

The model learns the flow $\vec{v}_t$ from the time derivative of the target interpolants: 
\begin{equation} \label{eq:target_interpolant}
    x_t := (1-t)\,x_0  + t \,x_1 
\end{equation}
with $x_0$ the $N_p$ initial test particles and $x_1$ the $N_p$ endpoints drawn on the target filaments. Fig.~\ref{fig:logigram} gives a schematic representation of the training method.
Considering that each filament is a curved line composed of several linear segments, we can recover the target filament structure $x_1$ by sampling at least three particles per segment. The value $N_p = 50,000$ is the lower bound of this condition and has been the choice for the following work.

Once the model is trained, it can be tested via the generation of filaments from a galaxy catalog and a set of initial test particles $x_0 \sim \nu_0$ : 
\begin{equation}
    \hat{x}_1 := x_0 + \int_0^1 dt \,\hat{v}_t
\end{equation}. 
Each iteration of $\hat{v_t}$ relies on the position of the test particles and the galaxies.

\begin{figure}[h!]
    \centering
    \begin{minipage}{.32\textwidth}
        \includegraphics[width=1.\linewidth]{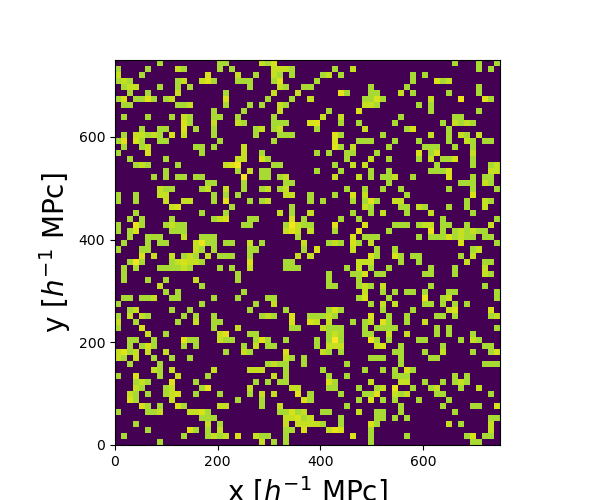}
    \end{minipage}
    \begin{minipage}{.32\textwidth}
        \includegraphics[width=1.\linewidth]{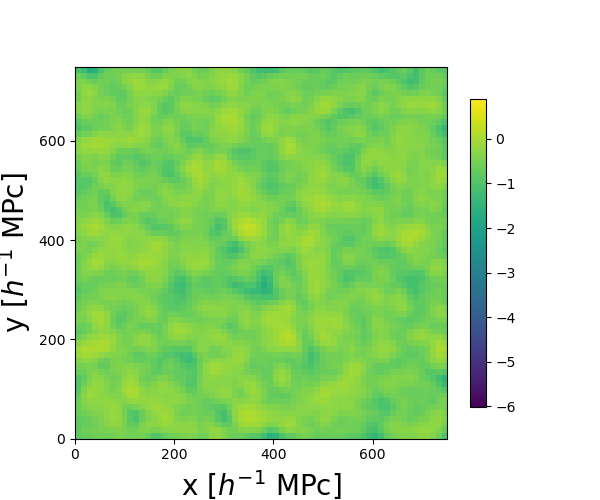}
    \end{minipage}
    \begin{minipage}{.27\textwidth}
  \centering
  \includegraphics[width=1.\linewidth]{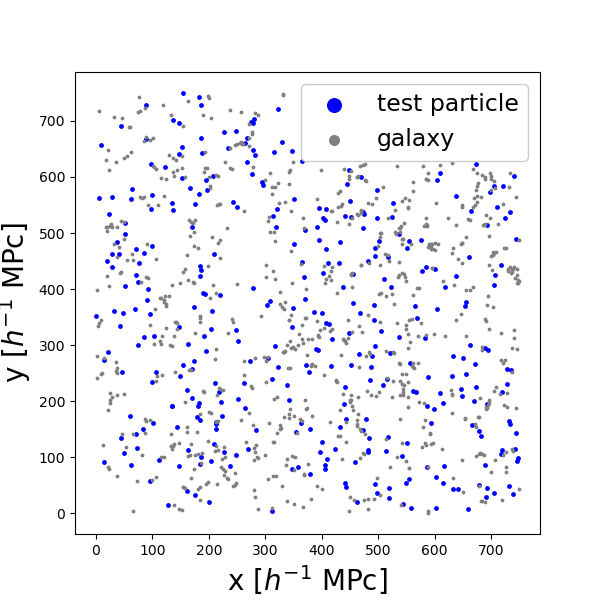}
\end{minipage}
    \caption{Distribution used for the initial sampling of test particles. \textit{Left:} given a galaxy distribution, we compute its density on a $64^3$ grid. \textit{Middle:} the density is smoothed with a Gaussian kernel whose standard deviation is set to $15\,h^{-1}\,\mathrm{Mpc}$. \textit{Right:} Sample from this probability density in the slice [85,100].}\label{fig:galaxy_sampling}
\end{figure}

In order to compute the $N_p$ target interpolants Eq.~\eqref{eq:target_interpolant}, we use ``data-dependent coupling''~\cite{Albergo2024} between the base and target densities.
We perform an efficient pair matching between $x_0$ and $x_1$ using the auction algorithm~\cite{Bertsekas1991} to approximately minimize the total distance particles have to traverse. 
Instead of choosing $\nu_0 =\mathcal{U} ([0 , L]^3)$, we introduce a bias in the initial density towards regions with high galaxy density.
Fig.~\ref{fig:galaxy_sampling} shows an example of such a sample. 

We use a Graph Neural Network (GNN)~\cite{Gori2005,Scarselli2009} to compute the particle velocities.
The input consists of a heterogeneous graph containing two types of nodes: test particles (P) and galaxies (G). The particles are connected to their $k_{PP}$ nearest neighbors
through particle-particle (PP) edges, while galaxies are linked to the $k_{GP}$ nearest particles through galaxy-particle
(GP) edges. Each node and edge carry a set of physical features describing the local environment
made up of particles and galaxies. The network then performs a message-passing process~\cite{Gilmer2017,Battaglia2018,Hamilton2018} on the graph and finally computes a flow.
We train the model using 8 Nvidia Tesla V100 GPUs (32GB). At the end of the project, 36 epochs have been completed in 720 hours, with the validation loss still improving.

\section{Results}


\begin{figure}
\centering
\begin{minipage}{.4\textwidth}
  \centering
\includegraphics[width=1.\linewidth]{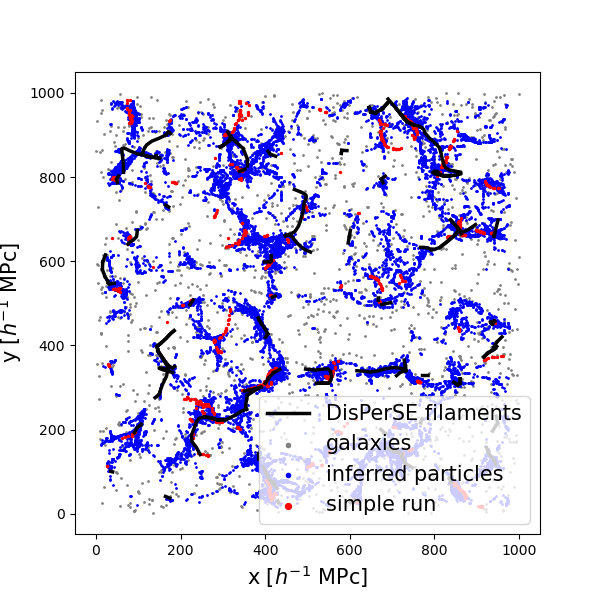}
\end{minipage}%
\begin{minipage}{.4\textwidth}
  \centering
\includegraphics[width=1.\linewidth]{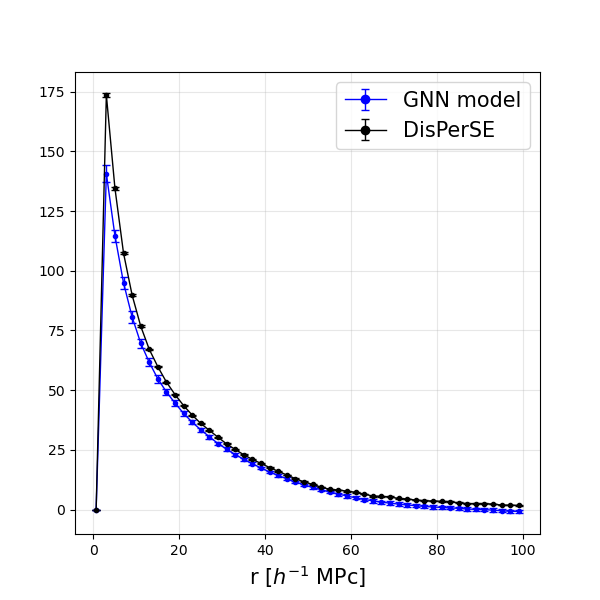}
\end{minipage}%
    \caption{\textit{Left:} Slice of $20\,h^{-1}\text{Mpc}$ thickness of the box where several runs with the trained model have been performed (in blue). A single run is highlighted in red.\textit{Right:} Average two-point correlation function of the test particles on 15 random initializations. The bars represent the deviation}
    \label{fig:inference_slice}
\end{figure}
We begin by presenting our model output in an example from the test set, displayed in the left panel of Fig.~\ref{fig:inference_slice}.
We integrate the model on 20 random initializations of the test particles to obtain an estimation of the probability density inferred by the network.
The DisPerSE filaments are mostly recovered while additional filaments are identified as well.
The full distribution in blue is closely linked to the probability inferred by the model. The main components shape a credible filament structure; one can notice the extra filaments appearing which can be originated from the stochasticity of the model or from an unlikely filament.
In red dots, a single sample is represented to give an impression of the sample variance.
We emphasize that the learned stochasticity is not a feature of the target as DisPerSE is a deterministic algorithm but rather a consequence of regularization due to finite training set and model capacity.
In the right panel of Fig.~\ref{fig:inference_slice} we plot the two-point correlation function of the test particles and the target particles.
The model has generally learned quite well what filaments look like.

\begin{figure}
\centering
\begin{minipage}{.4\textwidth}
  \centering
  \includegraphics[width=1.\linewidth]{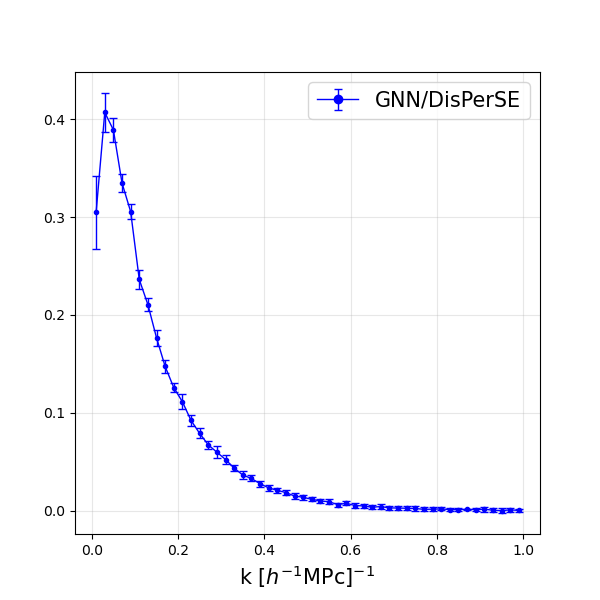}
\end{minipage}%
\begin{minipage}{.4\textwidth}
  \centering
  \includegraphics[width=1.\linewidth]{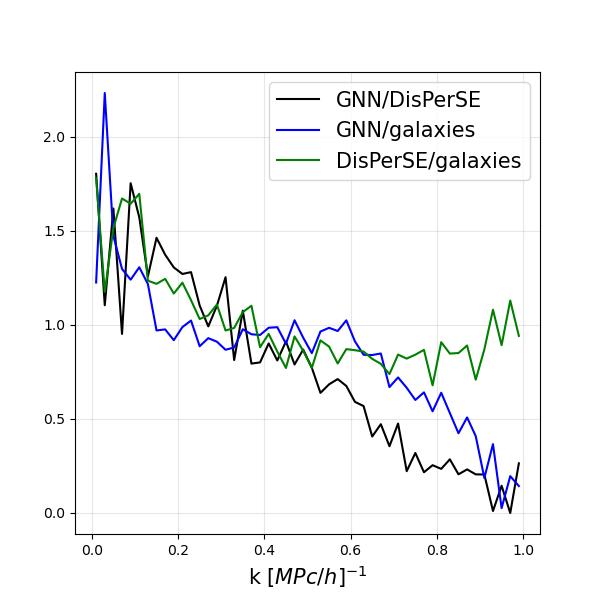}
\end{minipage}%
    \caption{\textit{Left:} Cross correlation between the test particles and the target particles with the same set up as in Fig~\ref{fig:inference_slice}. \textit{Right:} Signal $S := \frac{\text{Var}_{\text{close}}C - \text{Var}_{\text{random}}C}{\text{Var}_{\text{random}}C}$with $C$ being the cross-correlation for the test/target particles (black), the galaxies/test particles (blue), and the galaxies/target particles (green). 
    } 
    \label{fig:cross_correlation}
\end{figure}
In the left panel of Fig.~\ref{fig:cross_correlation}, we show the cross-correlation between test particles and target particles.
The correlation is moderate on large scales and decreases on smaller scales.
 We study the recovered filaments' sensitivity to cosmological parameters. To do so, we compute the cross-correlation between the galaxies and the particles for 15 random cosmological configurations and 15 close configurations (according to the values of $\Omega_m$ and $\sigma_8$). In the right panel of Fig.~\ref{fig:cross_correlation} the ratio of the two variances has been represented for the GNN model and for DisPerSE. 
According to this signal, the model's dependence on cosmological parameters is very close to DisPerSE's. 


\begin{SCfigure}[1]
  \includegraphics[width=0.4\linewidth]{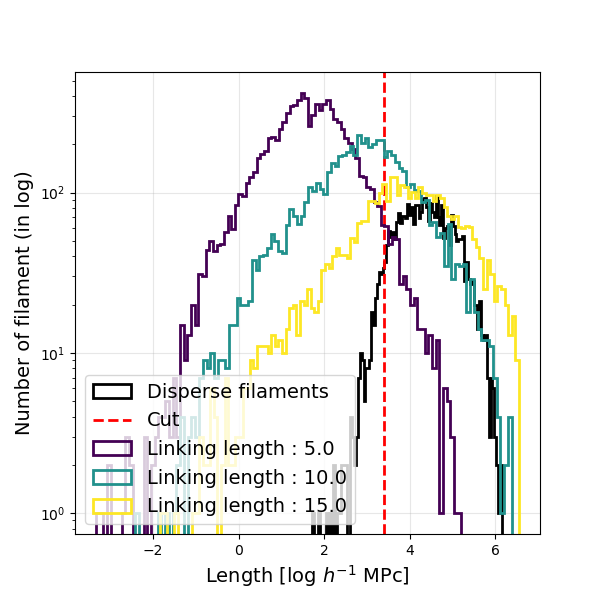}
  \caption{Histogram of filament lengths in log scale. In black the histogram of DisPerSE, and in other colors the histograms of the reconstruction algorithm for different values of $\beta$. In red dot, the line shows the cut introduced for the length of the filaments.}
  \label{fig:lengths}
\end{SCfigure}
In order to recover discrete filaments from the evolved test particles a linking procedure is required.
While we have not finalized this part of the algorithm (and might use an additional learning step for it), a quick solution is to use a clustering algorithm with maximum linking length $\beta$.
Fig.~\ref{fig:lengths} shows the length distribution of filament according to different values of $\beta$. The value of $\beta$ has been fixed according to the profile of 10 galaxy catalogs from the validation set. Compared to the DisPerSE teacher there is a large number of short filaments.
However, the distribution of long filaments is recovered well for $\beta=10\,h^{-1}\text{Mpc}$. The result is similar for all 10 catalogs.

\section{Discussion and conclusions}



We argue that constraining learned representations of galaxy catalogues with known cosmic web concepts is a promising path towards recovering more information from cosmological surveys. In this work, we take a first step towards linking galaxy catalogues with the underlying filamentary structure of the dark matter field. We train a generative model to evolve test particles according to a flow matching objective. The model reproduces summary statistics of the filaments well, despite differences with the teacher at the level of individual catalogs. We interpret these differences as the model finding additional filamentary structures, partly because the artificial filament-length cut imposed on the training data is difficult to learn. While this highlights a limitation of the current approach, the agreement in population-level statistics is an encouraging result.

Robust filament reconstruction is essential to exploit the scientific potential of the cosmic web in current and future galaxy surveys. Filament catalogs allow us to characterize the large-scale environments in which galaxies evolve, investigate the physical properties of filaments, and explore their potential as cosmological probes.

Our results represent a first step towards these scientific goals. Moving forward, we will construct a filament definition based on the underlying total matter density and velocity fields, moving beyond the deterministic DisPerSE targets used in this study. We will then train a graph flow to recover the stochastic mapping from observed galaxies to these filaments, and improve the method to link test particles into discrete filaments. Together, these developments aim to establish a framework for probabilistic cosmic web reconstruction, opening new possibilities for studying galaxy environments, filament physics, and cosmology.

\begin{ack}
A.T. acknowledges support from the International Laboratory for Astrophysics, Neutrino and Cosmology Experiments (ILANCE), a CNRS-University of Tokyo International Research Laboratory.
L.T. acknowledges support from JSPS Kakenhi grants 24K22878, 26K17136, JST PRESTO JPMJPR26287514, and Royal Society International Collaboration Award Round 2 (Japan) 252140.


\end{ack}


{
\small

\bibliography{main}
\bibliographystyle{unsrt}

}


\appendix

%


\end{document}